\documentclass[11pt]{article}
\pdfoutput=1
\usepackage[centertags]{amsmath}
\usepackage[square, comma, sort&compress,numbers]{natbib}
\usepackage{array,multirow}
\numberwithin{equation}{section}
\usepackage{amssymb}
\usepackage{graphicx}
\usepackage{color}
\usepackage[normalem]{ulem}

\usepackage{epsfig}
\usepackage{bbold}

\usepackage{wrapfig}
\usepackage{float}
\usepackage{soul}

\usepackage{tikz,pgf}
\usetikzlibrary{shapes}
\usetikzlibrary{calc}
\usetikzlibrary{decorations.pathmorphing}
\usetikzlibrary{decorations.pathreplacing,shapes.misc}
\usetikzlibrary{positioning}
\usetikzlibrary{arrows}
\usetikzlibrary{decorations.markings}
\usetikzlibrary{shadings}

\usetikzlibrary{intersections}

\def\sideremark#1{\ifvmode\leavevmode\fi\vadjust{\vbox to0pt{\vss
 \hbox to 0pt{\hskip\hsize\hskip1em
 \vbox{\hsize3cm\tiny\raggedright\pretolerance10000
  \noindent #1\hfill}\hss}\vbox to8pt{\vfil}\vss}}}

\def\be{\begin{equation}}
\def\ee{\end{equation}}
\def\ba{\begin{array}}
\def\ea{\end{array}}

\def\dps{\displaystyle}

\def\sd{D^\dagger}

\advance\textheight\topskip
\renewcommand{\tilde}{\widetilde}

\newtheorem{prop}{Proposition}[section]

\newcommand{\bref}[1]{\textbf{\ref{#1}}}

\newcommand{\im}{\mathop{\mathrm{Im}}}

\newcommand{\gh}[1]{\mathrm{gh}(#1)}

\newcommand{\dd}{\partial}

\renewcommand{\geq}{\,{\geqslant}\,}
\renewcommand{\leq}{\,{\leqslant}\,}

\newcommand{\binner}[2]{%
  {\langle}\kern-4.15pt{\langle}#1{,}\,#2{\rangle}\kern-4.15pt{\rangle}}

\newcommand{\half}{\mathchoice{%
    \ffrac{1}{2}}{\frac{1}{2}}{\frac{1}{2}}{\frac{1}{2}}}

\newcommand{\ffrac}[2]{\raisebox{.5pt}%
  {\footnotesize$\displaystyle\frac{#1}{#2}$}\kern1pt}

\newcommand{\brst}{\mathsf{\Omega}}

\renewcommand{\st}[2]{\overset{#1}{#2}}

\newcommand{\dl}[1]{\mathchoice{\ffrac{\dd}{\dd #1}}{\frac{\dd}{\dd
      #1}}{\ffrac{\dd}{\dd #1}}{\ffrac{\dd}{\dd #1}}}

\def\cE{\mathcal{E}}
\def\cF{\mathcal{F}}
\def\cG{\mathcal{G}}
\def\cH{\mathcal{H}}

\def\cN{\mathcal{N}}

\def\cQ{\mathcal{Q}}

\def\cT{\mathcal{T}}

\numberwithin{equation}{section} \makeatletter
\@addtoreset{equation}{section}

\def\be{\begin{equation}}
\def\ee{\end{equation}}
\def\ba{\begin{array}}
\def\ea{\end{array}}

\def\dps{\displaystyle}

\def\1{\tilde{1}}
\def\2{\tilde{2}}
\def\3{\tilde{3}}

\newdimen\tableauside\tableauside=1.0ex
\newdimen\tableaurule\tableaurule=0.4pt
\newdimen\tableaustep
\def\phantomhrule#1{\hbox{\vbox to0pt{\hrule height\tableaurule
width#1\vss}}}
\def\phantomvrule#1{\vbox{\hbox to0pt{\vrule width\tableaurule
height#1\hss}}}
\def\sqr{\vbox{%
  \phantomhrule\tableaustep

\hbox{\phantomvrule\tableaustep\kern\tableaustep\phantomvrule\tableaustep}%
  \hbox{\vbox{\phantomhrule\tableauside}\kern-\tableaurule}}}
\def\squares#1{\hbox{\count0=#1\noindent\loop\sqr
  \advance\count0 by-1 \ifnum\count0>0\repeat}}
\def\tableau#1{\vcenter{\offinterlineskip
  \tableaustep=\tableauside\advance\tableaustep by-\tableaurule
  \kern\normallineskip\hbox
    {\kern\normallineskip\vbox
      {\gettableau#1 0 }%
     \kern\normallineskip\kern\tableaurule}%
  \kern\normallineskip\kern\tableaurule}}
\def\gettableau#1 {\ifnum#1=0\let\next=\null\else
  \squares{#1}\let\next=\gettableau\fi\next}

\def\cE{\mathcal{E}}
\def\cF{\mathcal{F}}
\def\cG{\mathcal{G}}
\def\cH{\mathcal{H}}

\def\cN{\mathcal{N}}

\def\cQ{\mathcal{Q}}

\def\cT{\mathcal{T}}

\numberwithin{equation}{section} \makeatletter
\@addtoreset{equation}{section}

\def\be{\begin{equation}}
\def\ee{\end{equation}}
\def\ba{\begin{array}}
\def\ea{\end{array}}

\def\dps{\displaystyle}

\def\ba{\begin{array}}
\def\ea{\end{array}}

\def\dps{\displaystyle}

\usepackage{jheppub}

\makeatletter
\def\@fpheader{\vspace{-.1cm}}
\makeatother

\title{Continuous spin fields of mixed-symmetry type}

\author[a,b]{Konstantin\ Alkalaev,}

\author[a,c]{Maxim\ Grigoriev}

\affiliation[a]{I.E. Tamm Department of Theoretical Physics, \\P.N. Lebedev Physical
Institute,\\ Leninsky ave. 53, 119991 Moscow, Russia}
\affiliation[b]{Department of General and Applied Physics, \\
Moscow Institute of Physics and Technology, \\
Institutskiy per. 7, Dolgoprudnyi, \\141700 Moscow region, Russia}
\affiliation[c]{Arnold Sommerfeld Center for Theoretical Physics, \\Ludwig-Maximilians University Munich Theresienstr. 37, \\ D-80333 Munich, Germany}
\emailAdd{alkalaev@lpi.ru}
\emailAdd{grig@lpi.ru}

\abstract{ We propose a description of  continuous spin massless fields of mixed-symmetry type in Minkowski space at the level of equations of motion. It is based on the appropriately modified version of the constrained system originally used to describe massless bosonic fields of mixed-symmetry type. The description is shown to produce generalized versions of triplet, metric-like, and light-cone formulations. In particular, for scalar continuous spin fields we reproduce the  Bekaert-Mourad formulation and the Schuster-Toro formulation. Because a continuous spin system inevitably involves infinite number of fields, specification of the allowed class of field  configurations becomes a part of its definition. We show that the naive choice leads to an empty system and propose a suitable class resulting in the correct degrees of freedom. We also demonstrate that the gauge symmetries present in the formulation  are all Stueckelberg-like so that the continuous spin system is not a genuine gauge theory.

}

\preprint{FIAN-TD-2017-28}
\arxivnumber{}

\begin{document}

\maketitle
\flushbottom

\section{Introduction}

Continuous spin massless particles \cite{Bargmann:1948ck} have several interesting properties including the presence of a  dimensionful parameter $\mu$ and the  infinite number of physical degrees of freedom~\cite{Bekaert:2005in,Bengtsson:2013vra,Schuster:2014hca,Rivelles:2014fsa,Najafizadeh:2015uxa,Metsaev:2016lhs,Metsaev:2017ytk,Zinoviev:2017rnj,Najafizadeh:2017tin,Bekaert:2017khg,Rehren:2017xzn,Metsaev:2017cuz,Bekaert:2017xin,Khabarov:2017lth,Metsaev:2017myp}.  
From the conventional higher spin theory perspective \cite{Vasiliev:2003ev,Bekaert:2005vh} the most striking and intriguing feature is that the continuous spin dynamics can be defined on the space of fields which is the sum of Fronsdal-like  rank-$s$ fields with  $s = 0, ..., \infty$. The corresponding gauge invariant action functional  on Minkowski space $\mathbb{R}^{d-1,1}$ and $AdS_d$ space is the infinite sum of Fronsdal rank-$s$ actions with off-diagonal terms proportional to the continuous spin parameter $\mu$ \cite{Schuster:2014hca,Metsaev:2016lhs}. The gauge transformations are the standard Fronsdal transformations deformed by Stueckelberg-like terms also proportional to $\mu$. Moreover, the continuous spin fields can consistently interact with massive higher spin fields \cite{Metsaev:2017cuz,Bekaert:2017xin}.

From the group-theoretical perspective continuous spin particles correspond to infinite-dimensional massless UIRs of the Poincare algebra $iso(d-1,1)$, induced from infinite-dim\-ensional UIRs of $iso(d-2)$ subalgebra \cite{Bargmann:1948ck,Brink:2002zx,Bekaert:2006py}. The associated quantum numbers are the standard mass $m=0$, the continuous spin parameter $\mu\neq 0$, and (half-)integer spin numbers $(s_1, ..., s_p)$, where $p = [\frac{d-3}{2}]$. The description of \cite{Schuster:2014hca,Metsaev:2016lhs} was derived in the case of scalar representation. In the mixed-symmetry case the continuous spin dynamics on $\mathbb{R}^{d-1,1}$ was described at the level of equations of motion  \cite{Bekaert:2005in} as the particular contraction of the Fronsdal massless equations in $\mathbb{R}^{d,1}$ space.\footnote{For the frame-like Lagrangian formulation of continuous spin $(s_1,0,...,0)$ fields in $AdS_d$ see \cite{Khabarov:2017lth}, the light-cone dynamics of continuous spin $(s_1,0,...,0)$ fields  in $AdS_5$ space was considered in  \cite{Metsaev:2017myp}.}

In this paper  we study equations of motion for mixed-symmetry continuous spin fields in Minkowski space. To this end we use the generating formulation elaborated in  \cite{Barnich:2004cr,Alkalaev:2008gi,Alkalaev:2009vm,Alkalaev:2011zv} based on a first-quantized constrained system whose representation  space is interpreted as the space of field configurations.
We show that the continuous spin system corresponds to an appropriate deformation  of the constraints, which is  parameterized by the continuous spin parameter $\mu$. This allows us to formulate the triplet-like formulation of the continuous spin dynamics that generalizes the standard triplet formulation~\cite{Bengtsson:1986ys,Francia:2002pt,Sagnotti:2003qa,Alkalaev:2008gi}. Taking the triplet form as a starting point we derive the metric-like description which generalizes Fronsdal or Labastida formulations \cite{Fronsdal:1978rb,Labastida:1989kw}.\footnote{Free massless higher spin fields were discussed within various formulations, see e.g. \cite{Lopatin:1988hz,Buchbinder:2001bs,Zinoviev:2001dt,Alkalaev:2003qv,Skvortsov:2008sh,Campoleoni:2008jq,Skvortsov:2009zu,Boulanger2009,Campoleoni:2012th,Francia:2012rg,Bekaert:2015fwa,Joung:2016naf}.} Under the usual assumptions we also arrive at a light-cone formulation by eliminating quartets.

Because a continuous spin system inevitably involves infinite number of fields, specification of the allowed class of field  configurations becomes a part of its definition. Leaving aside space-time behavior of the fields we concentrate on the target space which is the space of ``functions'' in auxiliary oscillator variables. The simplest choice  which is polynomials in oscillators is not compatible with the deformed constraints in the sense that in contrast to the usual helicity spin case the constraints do not admit polynomial solutions. The way out would be to allow for formal series but as we demonstrate in this work with this choice gauge symmetry kills all the degrees of freedom. Nevertheless, it turns out that it is possible to identify a certain subspace of the formal series in which, on one hand, deformed constraints admit nontrivial solutions while, on the other hand, gauge symmetry does not kill everything. We then demonstrate by performing the light-cone analysis that the system indeed propagates correct degrees of freedom.

The paper is organized as follows. In Section \bref{sec:eom} we formulate the continuous spin dynamics 
in terms of the BRST first quantized system. To identify the continuous spin parameter we explicitly compute both the quadratic and  quartic Casimir operators of the Poincare algebra.  In Section  \bref{sec:triplet} we develop the triplet formulation of the continuous spin dynamics\footnote{BRST formulation of the continuous spin dynamics was previously discussed in \cite{Bengtsson:2013vra,Edgren:2006un}.}. Here we briefly review the homological reduction technique that we use to derive other forms of the continuous spin dynamics. In this section we also find the metric-like formulation which in the scalar continuous spin case reproduces the Bekaert-Mourad equations \cite{Bekaert:2005in} and the Schuster-Toro equations \cite{Schuster:2014hca} for an infinite collection of Fronsdal-like tensor fields of all ranks. In Section \bref{sec:light} we show how the triplet equations reduce to the light-cone equations. In particular, we explicitly describe the infinite-dimensional field space in terms of $o(d-2)$ tensors and calculate the Casimir operators of the Wigner  algebra $iso(d-2)$. In Section \bref{sec:module} we study the Weyl and gauge modules and show that both of them are trivial unless one assumes specific functional class in the sector of oscillator variables. Within this class the gauge module remains trivial while the Weyl module becomes non-trivial thereby indicating the system propagates physical degrees of freedom. Appendices \bref{sec:casimir}--\bref{app:shto} contain auxiliary and technical statements.

\section{Equations of motion for continuous spin fields }
\label{sec:eom}

In this section we formulate the continuous spin dynamics as viewed from the constraint algebra perspective.  The underlying constrained system is the modified version of that proposed in the helicity spin case \cite{Alkalaev:2008gi}, where some of the constraints are deformed by constant terms associated to the continuous spin parameter.

\subsection{Auxiliary variables and constraints}
\label{sec:cons}

A continuous spin massless system can be represented in terms of generating function $\phi(x,a)$, where $x^b$, $ b= 0, \ldots, d-1$ are Cartesian coordinates in Minkowski space $\mathbb{R}^{d-1,1}$ with the metric $\eta_{ab} = (+- \cdots-)$,  and $a_i^b$ are auxiliary commuting variables, $i = 1,\ldots,n$. 

The Poincare algebra $iso(d-1,1)$ basis elements are realized as 
\be
\label{poincare}
P_a = \dl{x^a}\;,
\qquad
M_{ab} = x_a \dl{x^b} - x_b\dl{x^a}+a_a{}_i \dl{a^b_i} - a_b{}_{i} \dl{a^a_i}\;.
\ee
Let us introduce notation
\be
\label{not}
\ba{c}
\dps
\Box = \frac{\partial^2}{\partial x^b\partial x_b}\;,
\qquad
\sd_i = a_i^b \frac{\partial}{\partial x^b}\;,
\qquad
D^i = \frac{\partial^2}{\partial a_i^b \partial x_b}\;,
\\
\\
\dps
T^{ij} = \frac{\partial^2}{\partial a_i{}_b \partial a_j^b}\;,
\qquad
T^\dagger_{ij}=a^b_i a_{bj}\;,\qquad 
N_i{}^j  = a_i^b \frac{\partial}{\partial a_j^{b}}\;, 
\qquad
N_i = a_i^b \frac{\partial}{\partial a_i^{b}}\;.
\ea
\ee  
The above operators form a subalgebra in $sp(2n+2)$ algebra which is dual to the Lorentz algebra formed by generators $M_{ab}$ in the sense of Howe \cite{Howe}. The space of formal series in $a^b_I = (x^b, a_i^b)$ is an $iso(d-1,1)\oplus sp(2n+2)$ bimodule. This algebraic framework was employed in~\cite{Alkalaev:2008gi} to describe usual mixed-symmetry gauge fields.

The continuous spin system is described in terms of the (suitably modified) constraints from \eqref{not}. Moreover, all the constraints but $D^\dagger_i$ are imposed directly while constraints $D^\dagger_i$ are imposed in a dual way as generators of gauge transformations.


\paragraph{Differential constraints.} These are the same as in the helicity spin case, 
\be
\label{diff}
\Box \phi = 0\;, \qquad D^i \phi = 0\;,
\qquad
i=1,...,n\;.
\ee

\vspace{-3mm}

\paragraph{Algebraic constraints.} We impose the modified trace and Young symmetry constraints 
\be
\label{Tij}
(T^{ij}+\nu^{ij})\phi = 0\;,
\qquad
\nu^{ij} = \nu \,\delta^{1i}\delta^{1j}\;,
\qquad 
i,j = 1,\ldots,n\;,
\ee
\be
\label{young}
N_i{}^j \phi = 0\; \quad i<j\;,
\qquad
N_i\phi =  s_i \phi\;,\qquad i,j = 2,\ldots,n\;, 
\ee
where the spin weights $s_i\geq 0$ are non-negative integers and  $\nu \in \mathbb{R}$. 

\vspace{-3mm}

\paragraph{Gauge equivalence.} The gauge transformations are given by 
\be
\label{mui}
\delta \phi  = \left(\sd_i+\mu_i\right) \chi^i\;, 
\qquad
\mu_i = \mu \,\delta_{1i}\;, \qquad 
i = 1,\ldots,n\;,
\ee
where $\chi^i$ are the gauge parameters satisfying the off-shell constraints that follow from \eqref{diff}--\eqref{young} and $\mu \in \mathbb{R}$.

We note that the constraints are consistent provided that  the parameter matrices $\nu^{ij}$ and $\mu^i$ are fixed as in \eqref{Tij} and \eqref{mui}. In Section \bref{sec:C4} we show that $\nu\mu^2$ is  the value of the quartic Casimir operator of the Poincare algebra $iso(d-1,1)$ so that fixing e.g. $\nu = 1$ we find out that $\mu$ is the continuous spin parameter. Note also that a similar constraint system was discussed in~\cite{Bekaert:2005in}.

In contrast to the constrained system describing the helicity spin fields the employed constraints are not the highest weight conditions of  $sp(2n+2)$ algebra. On the contrary, the deformed constraints are typical for the theory of coherent states, where the states are defined as eigenstates of the annihilation operator (e.g., $T^{ij}$ in our case). It follows that such elements do not diagonalize the particle number operator anymore (missing  $N_1$ in our case) and are represented as infinite power series in auxiliary variables.

As the formulation involves operators that mix tensors of different rank the system non-trivially involves infinite number of fields and, hence, it is important to specify which functions in $a^b_i$ are allowed because this defines the field content of the theory. The choice to be motivated later (see Section~\bref{sec:modules}) is as follows: we take formal series in $a^b_i$ satisfying the additional admissibility condition. A series $f$ is admissible if its trace decomposition 
\begin{equation}
\label{func-class-exp}
 f=f_0+f_1^{ij}T^\dagger_{ij}+f_2^{ij,kl}T^\dagger_{ij}T^\dagger_{kl}+\ldots \,, \qquad T^{ij}f^{\dots}_p=0\,,
\end{equation} 
is such that all coefficients are polynomials of finite order (i.e. for a given $f$ there exists such $N\in \mathbb{N}$ that all $f_r$ are of order not exceeding $N$). It is clear that both Poincare and $sp(2n+2)$ algebras preserve the space of admissible elements. This space serves as the target space of the system.

\subsection{BRST operator}
\label{sec:brst_mixed}

In what follows we actively employ BRST first quantized formalism, see e.g.~\cite{Barnich:2004cr,Alkalaev:2008gi} for more details and original references. To this end, we introduce  the anticommuting ghost  variable $b^i$ with the ghost number  $\gh{b_i} = -1$, and then split both the auxiliary commuting variables and ghost variables as  $a^b_i = (a^b, a^b_\alpha)$ and $b^i = (b, b^\alpha)$, where $\alpha = 1,...,n-1$. Fields of the system as well as (higher-order) gauge parameters are encoded in the generating function $\Psi(x,a|b)$. It can be expanded in $b_i$ so that  homogenous components have definite ghost degrees, i.e. 
\be
\label{decgh}
\Psi = \sum_{n\geq 0} \Psi^{(-n)}\;, \qquad  \gh{\Psi^{(-n)}} = -n\;.
\ee
According to the usual prescription fields are encoded in degree $0$ component, gauge parameters in degree $-1$ component, etc.  

The system \eqref{diff}--\eqref{mui} can now be written in the BRST form. All constraints remain the same except for the Young symmetry and spin  conditions \eqref{young} that receive ghost extensions. Namely, 
\be
\label{ext1}
\cN_\alpha{}^\beta \Psi = 0 \quad \alpha<\beta\;,  \qquad \cN_\alpha\Psi = s_\alpha\Psi\;,
\qquad \alpha,\beta = 1, ...,n-1\;,
\ee
where 
\be
\cN_\alpha{}^\beta = N_\alpha{}^\beta + b_\alpha\dl{b_\beta}
\qquad \text{and}\qquad 
\cN_\alpha = N_\alpha+ b_\alpha\dl{b_\alpha}\;,
\ee
and no sum over $\alpha$ in the last formula is implied. In this way the above constraints simultaneously impose the conditions on fields as well as on gauge parameters.

The BRST operator is given by   
\be
\label{gauge}
\cQ = \left(\sd_i+\mu_i\right) \dl{b_i} \equiv \left(\sd+\mu\right) \dl{b} + \sd_\alpha\dl{b_\alpha}\;,
\qquad
Q^2 = 0\;.
\ee
The BRST invariance of the constraints fixes parameters $\nu^{ij}$ and $\mu^{i}$ to be proportional to arbitrary $\nu$ and $\mu$.\footnote{To this end, we compute  
$[Q,\cN_i{}^j ] = \mu_i\dl{b^j}$ and 
$[T^{mn},\cN_i{}^j] = \delta_i^n T^{mj}+ \delta_i^m T^{nj}$ and imposing the constraints we find out that the parameters must be as in \eqref{Tij} and \eqref{mui}.} To reproduce the gauge transformation \eqref{mui} we identify fields $\phi(x,a)$ as ghost number zero components $\phi(x,a) = \Psi^{(0)}(x,a)$ and gauge parameters $\chi^i(x,a)$ as ghost minus one components $b_i\chi^i(x,a) = \Psi^{(-1)}(x,a|b)$, cf. \eqref{decgh}. Then, the gauge transformation $\delta \Psi^{(0)} = \cQ \Psi^{(-1)}$ yields  \eqref{mui}, where parameters $\chi^i(x,a)$ satisfy the same constraints \eqref{diff}, \eqref{Tij}, while \eqref{young} are appropriately modified.

\subsection{Evaluating the Casimir operators}
\label{sec:C4}

Our formulation involves parameters $\mu,\nu$ and $(n-1)$ spin weights $s_1, ..., s_{n-1}$. In $d$ dimensions $n =[\frac{d-3}{2}]$  that allows describing all possible finite-dimensional modules of the short little algebra $o(d-3)\subset iso(d-1,1)$ \cite{Brink:2002zx}. 
 
To characterize  $iso(d-1,1)$ representations underlying the system \eqref{diff}-\eqref{mui} we analyze the Casimir operators of the Poincare algebra  \eqref{poincare} briefly reviewed in Appendix \bref{sec:casimir}. The quadratic Casimir operator  $C_2 = P_aP^a \approx 0 $ vanishes on-shell because of the constraint \eqref{diff}. Then, the 
quartic Casimir operator $C_4  = (M_{ab}P^b)^2$ equals    
\be
\label{onshellcas}
C_{4}\,\phi(x,a) =  - \sd_i \sd_j \,T^{ij} \,\phi(x,a) \approx \; \mu^2 \nu\,  \phi(x,a)\;, 
\ee
where we used the  differential constraints \eqref{diff}, trace constrains \eqref{Tij} and the equivalence relation  $\phi \sim \phi + \cQ \chi$ with  the gauge parameter expressed in terms of the field $\phi$.\footnote{The analogous consideration of the Casimir operators on the equivalence classes in the case of $o(d-1,2)$ algebra and $AdS_d$ gauge fields can be found in \cite{Alkalaev:2011zv}.}  From our analysis in Appendix \bref{sec:casimir} it follows that any higher order  Casimir operator is also proportional to $ \mu^2 \nu$. 

Thus, we see that the model propagates continuous spin particles, in which case fixing $\nu=1$ we identify $\mu$ as the continuous spin parameter. On the other hand, we stress that such a split between deformation parameters $\mu$ and $\nu$ is artificial and only their combination
$\mu^2 \nu$ has invariant meaning.

\section{Triplet  formulation}
\label{sec:triplet}

Now we extend the triplet formulation of  helicity spin fields of symmetric and mixed-symmetry type \cite{Bengtsson:1986ys,Francia:2002pt,Sagnotti:2003qa,Alkalaev:2008gi} to the continuous spin case. To this end, we introduce additional anti-commuting ghost variables $c_0, c_i$, $i = 1,\ldots,n$, with positive ghost numbers $\gh{c_0} = 1$, $\gh{c_i} = 1$. These variables are associated with the differential constraints \eqref{diff}. 

Then, the BRST operator  \eqref{gauge} can be extended as follows
\begin{equation}
\label{standard-brst}
\brst=c_0\Box +c_i D^i+\big(D^{\dagger}_ i+\mu_i\big) \dl{b_i}-c_i\dl{b_i}\dl{c_0}\,, 
\end{equation}
where $\mu_i = \mu \delta_{i1}$. It is defined on the subspace of  $\Psi = \Psi(x,a|c,b)$ singled out by the BRST extended trace constraints
\be
\label{BRSTtr}
(\cT+\nu)\Psi = 0\;, 
\qquad
\cT^{\alpha}\Psi = 0\;,
\qquad
\cT^{\alpha\beta}\Psi = 0\;,
\ee
as well as the Young symmetry and the spin weight constraints 
\be
\label{ext}
\cN_\alpha{}^\beta \Psi = 0\quad \alpha<\beta\;,  \qquad \cN_\alpha\Psi = s_\alpha\Psi\;.
\ee
The extended constraints read explicitly as
\be
\label{conex}
\ba{c}
\dps
\cT^{ij} = T^{ij}+ \dl{c_i}\dl{b_j}+\dl{c_j}\dl{b_i}\;,
\qquad
\cN_\alpha{}^\beta = N_\alpha{}^\beta +b_\alpha\dl{b_{\beta}} + c_\alpha\dl{c_{\beta}}\;,
\\
\\
\dps
\cN_\alpha = N_\alpha+ b_\alpha\dl{b_\alpha}+ c_\alpha\dl{c_\alpha}\;,
\qquad
\alpha,\beta = 1, ...,n-1\;,
\ea
\ee
where in the last expression no summation over $\alpha$ is implied. Note that the BRST operator~\eqref{standard-brst} is nilpotent on the entire space of unconstrained fields and not only on the subspace singled out by~\eqref{BRSTtr} and \eqref{ext}. On the entire space BRST operator \eqref{standard-brst} describes a reducible system whose analog in the helicity case is well known in the literature~\cite{Bengtsson:1986ys,Francia:2002pt,Sagnotti:2003qa} and is relevant in the context of tensionless strings.

Expanding $\Psi$ into homogeneous components of definite ghost degree we concentrate on the vanishing degree component $\Psi^{(0)}$.
Representing then $\Psi^{(0)}$ as $\Psi^{(0)}  = \Phi + c_0 C$ we introduce component fields entering $\Phi = \Phi(x,a|b,c)$ and $C =C(x,a|b,c)$ according to 
\be
\label{PhiC}
\ba{l}
\dps
\Phi= \sum_{k=0}^{n-1}c_{i_1} ... c_{i_k} b_{j_1} ... b_{j_k} \Phi^{i_1 ... i_k|j_1 ... j_k}\;,
\qquad 
C= \sum_{k=0}^{n-2}c_{i_1} ... c_{i_k} b_{j_1} ... b_{j_{k+1}} C^{i_1 ... i_k|j_1 ... j_{k+1}}\;.
\ea
\ee
These component fields can be identified as generalized triplet fields \cite{Bengtsson:1986ys,Sagnotti:2003qa,Alkalaev:2008gi}. The corresponding gauge transformation reads
\be
\label{gt}
\delta \Psi^{(0)} = \brst \Psi^{(-1)}\;,
\ee
where the ghost number $-1$ parameters $\Psi^{(-1)} = \Lambda + c_0 \Upsilon$ are given by 
\be
\label{LU}
\ba{l}
\dps
\Lambda= \sum_{k=0}^{n-2}c_{i_1} ... c_{i_k} b_{j_1} ... b_{j_{k+1}} \Lambda^{i_1 ... i_k|j_1 ... j_{k+1}}\;,
\qquad \Upsilon= \sum_{k=0}^{n-3}c_{i_1} ... c_{i_k} b_{j_1} ... b_{j_{k+2}} \Upsilon^{i_1 ... i_k|j_1 ... j_{k+2}}\;.
\ea
\ee
Analogously, the ghost number $-k$ component $\Psi^{(k)}$ encodes $(k-1)$-th level reducibility parameters. 

Finally, the triplet equations of motion for continuous spin fields have the form  
\be
\label{eomtriplet}
\brst \Psi^{(0)} = 0\;.
\ee
By construction,   \eqref{eomtriplet} is invariant with respect to the gauge transformation \eqref{gt}, where both fields and parameters are constrained by \eqref{BRSTtr}--\eqref{ext}. We note that the BRST operator~\eqref{standard-brst} for the continuous spin system   differs from the BRST operator for the helicity spin system \cite{Alkalaev:2008gi} by adding the term proportional to $\mu$, i.e. $\brst \rightarrow \brst+\mu\dl{b}$.

\subsection{Homological reduction }   
\label{sec:red}

In this section we shortly review the homological reduction that can be applied to any gauge system defined by a BRST operator \cite{Barnich:2004cr}\footnote{In the context of the unfolded formulation of higher spin fields the homological technique to identify auxiliary fields and Stueckelberg variables was proposed in~\cite{Lopatin:1988hz,Shaynkman:2000ts}}. In the formal language the linear gauge system can be defined as a pair  $(\cH, \Omega)$, where $\cH$ is the representation space of the BRST first-quantized system and $\Omega$ is the nilpotent BRST operator. In addition, it is assumed that $\cH$ is graded by ghost degree in such a way that $\Omega$ carries degree $1$. Ghost degree zero elements $\Psi^{(0)}$ of $\cH$ are identified with field configurations while those at negative ghost degree with  (higher order) gauge parameters.  As before, the equations of motion are $\Omega\Psi^{(0)} = 0$ while gauge transformation are  $\delta\Psi^{(0)}=\Omega \Psi^{(-1)}$. Similarly one defines higher order gauge transformations (also known as reducibility relations).

Suppose $\cH$ can be split into three subspaces: $\cH = \cE\oplus \cF \oplus \cG$.
Let $\st{\cG\cF}{\Omega}:\cF \to \cG$ denotes $\Omega$ restricted to $\cF$ and projected to $\cG$, i.e. $\st{\cG\cF}{\Omega}f=( \Omega f)|_{\cG}$, where $f\in \cF$. If $\st{\cG\cF}{\Omega}:\cF \to \cG$ is invertible then all the fields associated with $\cF$ and $\cG$ are generalized auxiliary fields. The notion  of generalized auxiliary fields was introduced in~\cite{Henneaux:1990ua} and extended to the case of not necessarily Lagrangian systems in \cite{Barnich:2004cr}. In particular, among generalized auxiliary fields one finds usual auxiliary fields and Stueckelberg fields as well as the associated ghosts and antifields. 

Generalized auxiliary fields can be eliminated, resulting in a new gauge theory or, rather, another formulation of the same theory.  Theories related through elimination/addition  of generalized auxiliary fields are considered equivalent. Note that typically one is interested in local gauge field theories in which case one requires that generalized auxiliary fields can be eliminated algebraically. In the present case this corresponds to requiring $\st{\cG\cF}{\Omega}$ to be algebraic. Eliminating generalized auxiliary fields associated to $\cF,\cG$ gives a reduced theory $(\cE, \tilde \Omega)$, where $\cE$ is the representation space of the reduced BRST operator $\tilde\Omega$ given by
\begin{equation}
 \tilde\Omega
  =(\st{\cE\cE}{\Omega}
  - \st{\cE\cF}{\Omega}(\st{\cG\cF}{\Omega})^{-1}
  \st{\cG\cE}{\Omega})\,.
  \end{equation}

It is often useful to identify $\cE$ as a cohomology of a certain piece of $\Omega$. More specificaly,
suppose that $\cH$ admits an additional  grading 
\be
\label{hom}
\cH = \bigoplus_{-N}^\infty \cH_i\,, 
\ee
with  $N$ finite integer and such that $\Omega$ decomposes into homogeneous componets as follows
\be
\Omega =\Omega_{-1}+ \Omega_0+ \Omega_{1}+ \ldots\;. 
\ee  
It follows that the lowest grade part of the BRST operator $\Omega_{-1}$ defines the decomposition introduced above  
\be
\cE\oplus \cG = {\rm Ker}\, \Omega_{-1}\;,
\qquad
\cG = \im \Omega_{-1}\;,
\qquad
\cE = \frac{{\rm Ker}\, \Omega_{-1}}{\im \Omega_{-1}}\equiv H(\Omega_{-1})\;.
\ee
Note that by construction such $\st{\cG\cF}{\Omega}$ is invertible. 

In what follows the homological reduction technique is applied to find equivalent forms of the triplet formulation \eqref{standard-brst}: the metric-like formulation that generalizes  Fronsdal and Labastida formulations to the continuous spin case, and the light-cone formulation.  
   
\subsection{Metric-like formulation}

Let the additional grading \eqref{hom} be a homogeneity degree in $c_0$. Then, BRST operator  \eqref{standard-brst} can be decomposed as $\brst=\brst_{-1}+\brst_{0}+\brst_{1}$ 
with
\be
\label{3brst}
\brst_{-1}=-c_i\dl{b_i}\dl{c_0}\,,
\qquad
\brst_0=c_i D^i+\big(D^{\dagger}_ i+\mu_i\big) \dl{b_i}\,,
\qquad
\brst_1=c_0\Box\;.
\ee
The theory can be consistently reduced to the subspace $\cE = H(\brst_{-1})$ using the homological technique reviewed in Section \bref{sec:red}. Note that the $\mu$-deformation term enters only $\brst_{0}$ and, therefore, the cohomology $H(\brst_{-1})$ remains the same as in the helicity spin case \cite{Alkalaev:2008gi}.

The cohomology $H(\brst_{-1})$ in ghost degree $0$ and $-1$ can be explicitly described in terms of the lowest expansion  components in ghosts $c_i$ and $b^i$. More precisely, denoting the lowest component in \eqref{PhiC} as $\varphi$ we find $\Phi  = \varphi + \ldots\;$, where the ellipses denote the ghost contributions expressed in terms of traces of $\varphi$. Analogously, we denote by $\chi^i$ the lowest component of the gauge parameter $\Lambda^i$ in  \eqref{LU}. 
From \eqref{BRSTtr}--\eqref{conex} it then follows that $\varphi$ and $\chi^i$ satisfy the modified trace conditions 
\be
\label{beg1}
\mathbb{T}^{(ij}\mathbb{T}^{kl)}\varphi = 0\;, 
\qquad
\mathbb{T}^{(ij}\chi^{k)} = 0\;,
\ee
where we introduced the notation $\mathbb{T}^{ij} \equiv T^{ij}+\nu \,\delta^{i1}\delta^{j1}$. Young symmetry and spin weight conditions take then the form  
\be
\label{end0}
N_\alpha{}^\beta \varphi =0 \quad \text{at}\quad \alpha < \beta\qquad \text{and}\qquad  N_\alpha \varphi  =  s_\alpha \varphi\;,
\ee
and
\be
\label{end1}
N_\alpha{}^\beta \chi^\gamma + \delta_\alpha^\gamma  \chi^\beta =0\quad \text{at}\quad \alpha <\beta \qquad \text{and}  
\qquad 
N_\alpha\chi  = s_\alpha \chi\;,
\quad
N_\alpha \chi^\alpha = (s_\alpha-1)\chi^\alpha\;,
\ee
where $N_\alpha{}^\beta$ and $N_\alpha$ are given by \eqref{not} and no sum over $\alpha$ in the last equation of \eqref{end1} is implied. Let us note that the Young and weight conditions are imposed in the sector of $a^b_\alpha$ variables only. However, there are cross-trace conditions in \eqref{beg1} that mix up expansion coefficients in $a^b$ and $a^b_\alpha$.   

Now, using the above cohomological results we can  straightforwardly find the reduced equations of motion. Introducing operator $Z$ via $\brst_{-1} \equiv - \dl{c_0} Z$ the original triplet equations \eqref{eomtriplet} can be cast into the form
\be
\label{thursday26}
\Box \Phi  - \brst_{0}C = 0\;,
\qquad 
\brst_{0}\Phi - Z C = 0\;,
\ee
where fields $\Phi$ and $C$ are defined by the expansion  \eqref{PhiC}. It follows from the stucture of $H(\brst_{-1})$ that $C$ is an auxiliary field and, therefore,  using the second equation in \eqref{thursday26} it can be expressed in terms of $\brst_0 \Phi$. In other words, $C$ is given by derivatives of $\Phi$, while $\Phi$ itself is reduced to the lowest component $\varphi$. It follows~\eqref{thursday26} take the form
\be
\Box \varphi - (D^\dagger_i+\mu_i)C^i = 0\;,
\qquad
D^i \varphi  - (D^\dagger_j+\mu_j)\Phi^{i|j}-C^i=0\;,
\ee
where the component $\Phi^{i|j}$ can be expressed via $\varphi$ by virtue of the trace conditions \eqref{BRSTtr} as $\Phi^{i|j} = \half \mathbb{T}^{ij} \varphi$. Eliminating the auxiliary field $C^i$ we finally arrive at the reduced equations of motion
\be
\label{lab}
\left[\Box -(D_i^{\dagger}+\mu_i)D^i  + \half(D_i^{\dagger}+\mu_i)(D_j^{\dagger}+\mu_j) (T^{ij}+\nu^{ij})\right]\varphi   = 0\;, 
\ee
which are invariant with respect to the gauge transformations
\be
\delta \varphi = (D_i^{\dagger}+\mu_i)\chi^i\;.
\ee
Here, fields and gauge parameters are subject to the algebraic conditions \eqref{beg1}--\eqref{end1}. Note that setting $\mu, \nu =0$ we reproduce the  Labastida formulation \cite{Labastida:1989kw}.

\subsection{Scalar continuous spin case}
\label{sec:shto}

There is an interesting  formulation of the continuous spin dynamics due to Schuster and Toro, where the field content is that of infinite sum of Fronsdal systems while the equations and gauge symmetries involve off-diagonal terms proportional to the parameter $\mu$~\cite{Schuster:2014hca}. In this section we show that the Schuster-Toro equations can be obtained from the metric-like equations of the previous section  by choosing $n=1$. In this case the spin weights vanish $s_i=0$, $i=1,2,\ldots$ so that we deal with the scalar continuous spin system. 

The reduced equations of motion \eqref{lab}  take the form
\be
\label{red_eom}
\Box \varphi -(D^{\dagger}+\mu)D \varphi + \half(D^{\dagger}+\mu)^2 (T+\nu)\varphi   = 0\;.
\ee
These equations were proposed by Bekaert and Mourad in~\cite{Bekaert:2005in}. By construction, these are invariant under transformations  
\be
\label{deform}
\delta \varphi  = D^{\dagger}\epsilon  + \mu \epsilon\;,
\ee
supplemented with the deformed trace conditions
\be
(T+\nu)^2 \varphi = 0\;,\qquad (T+\nu)\epsilon = 0\;.
\ee
Note that there are no spin weight conditions in this case. However, the dynamics cannot be restricted to the spin-$s$ subspace since the deformed trace constraints are incompatible with the spin-$s$ weight condition $N\phi = s \phi$. Nonetheless, sending both $\nu$ and $\mu$ to zero we reproduce a direct sum of the Fronsdal equations for all integer spins.  
 
The deformed trace conditions can be explicitly solved in terms of tensors subjected to the standard trace conditions. Namely, in Appendix \bref{app:A} we demonstrate that fields and parameters can be equivalently represented as
\be
\label{trd}
\varphi = \sum_{n,m=0}^\infty \beta_{m,n} (T^\dagger)^m \varphi_{(n)}\;, \qquad 
\epsilon = \sum_{n,m=0}^\infty \beta_{m,n+1} (T^\dagger)^m \epsilon_{(n)}\;,
\ee
where $T^\dagger  = a^b a_{b}$ is the trace creation operator, the rank-$n$ tensors on the right-hand sides satisfy the Fronsdal conditions 
\be
\label{fr}
T^2 \varphi_{(n)} = 0\;,
\qquad 
T \epsilon_{(n)} = 0\;,
\ee 
while the coefficients $\beta_{m,n}$ are explicitly  given by \eqref{F2}. We conclude that  original $\varphi$ and  $\epsilon$ are replaced now by infinite collections of Fronsdal (single and double traceless) tensors of ranks running from zero to infinity.  

In Appendix \bref{app:shto} we explicitly show that in the Fronsdal basis \eqref{fr} the metric-like equations \eqref{red_eom} take the Schuster-Toro form \cite{Schuster:2014hca,Metsaev:2016lhs}
\be
\label{shto}
-\Box \varphi_{(n)} + D^\dagger G_{(n-1)} + \mu \left[G_{(n)} +  d_n\, T^\dagger G_{(n-2)}\right] = 0\;,\qquad n= 0,1,2,...\;.
\ee
Here, 
\be
\label{Gn}
G_{(n)} =  A_{(n)}+ \mu \,c_n B_{(n)} \;,
\ee
with the derivative and algebraic terms combined into  
\be
\label{AB}
\ba{l}
\dps
A_{(n)} =  D \varphi_{(n+1)} - \half D^\dagger\, T \varphi_{(n+1)}\;,
\qquad
B_{(n)} = \varphi_{(n)} + a_n\, T^\dagger T  \varphi_{(n)}+ b_n\, T \varphi_{(n+2)}\;,
\ea
\ee
where the coefficients are given by 
\be
\begin{gathered}
a_n =  - \frac{1}{2d+2n-8}\,,
\qquad
b_n = \frac{d+2n-2}{2\nu}\,,
\\
c_n =  - \frac{1}{2b_n}\,,
\qquad
d_n = -\frac{\nu}{(d+2n-4)(d+2n-6)}\,.
 \end{gathered}
\ee
We note that  $A_{(n)}$ and $B_{(n)}$ as well as $G_{(n)}$ are traceless. These combinations of fields and their derivatives are convenient to build the double-traceless operator \eqref{shto}.  

The gauge transformation \eqref{deform} reads 
\be
\label{shto_trans}
\delta \varphi_{(n)} = D^\dagger\epsilon_{(n-1)}+ \mu\,\left[ \epsilon_{(n)} + d_n \,T^\dagger \epsilon_{(n-2)}\right]
\;.
\ee 
This is the Stueckelberg-like transformation law with three different rank traceless gauge parameters, which is  typical for massive higher spin theories \cite{Zinoviev:2001dt}.

\section{Light-cone formulation for continuous spin fields}
\label{sec:light}

To formulate the light-cone dynamics we start from the triplet formulation and eliminate unphysical degrees of freedom by means of the homological reduction. 
This is achieved by using the suitable grading \cite{Aisaka:2004ga} (see also~\cite{Barnich:2005ga,Alkalaev:2008gi}) such that the so-called quartets form  the subspace $\cF\oplus \cG$ while the cohomology of the lowest degree piece of the BRST operator form the complementary subspace $\cE$ describing configurations of $o(d-2)$ tensor fields on the light-cone. 

The quartet grading is defined by\footnote{The light-cone coordinates are $x^{\pm} = \frac{1}{\sqrt{2}}(x^0\pm x^{d-1})$ and $x^m$, where $m=1,...,d-2$. $o(d-2)$ indices are denoted by letters from the middle of the alphabet $m,n,k,l,p,s,... $.  The scalar product reads $A_c B^c = A^+ B^-+A^-B^+-A^m B^m$. }
\be
\deg{a_i^{\pm}} = \pm 2\;,
\qquad
\deg{a_i^m} = 0\;,
\qquad 
\deg{c_0} = 0\;,
\qquad
\deg{c_i} = 1\;,
\qquad
\deg{b^i} = -1\;.
\ee
Note that in the assumed functional space this grading is bounded from both above and below because $\deg{T^\dagger}=0$ so that only the coefficients of the series~\eqref{func-class-exp} may contribute to the degree of an element. By the assumption for a given element these coefficients are polynomials of finite order in $a$-oscillators and hence the degree is finite. Thus, the conditions for the applicability of the homological reduction technique are satisfied. 

The triplet  BRST operator \eqref{standard-brst} decomposes into the homogeneous degree components as $\brst=\brst_{-1}+\brst_{0}+\brst_{1}+\brst_{2}+\brst_{3}$, where
\be
\label{quarta} 
\ba{c}
\dps
\brst_{-1} = p^+\left(c_i\dl{a_i^+}+a_i^{-}\dl{b_i}\right)\;, 
\qquad
\brst_{0} = c_0 (2 p^+ p^- + p_m p^m) \;,
\\
\\
\dps
\brst_{1} = c_i p^m \dl{a^m_i}+ p^+ a_i^- \dl{b_i} + \mu \dl{b}\;,
\quad
\brst_{2} = -c_i\dl{b_i}\dl{c_0}\;,
\quad 
\brst_{3} = p^-(c_i \dl{a^-_i} + a^+_i \dl{b_i})\;.  
\ea
\ee

We note that  $\brst_{-1}$ is the de Rham differential in variables $c_i, b_i$ and $a_i^-, a_i^+$. Thus, $H(\brst_{-1})$ identified with dynamical fields consists of elements depending on  transverse oscillators only  $\phi = \phi(x|a_i^m)$ \cite{Barnich:2005ga,Alkalaev:2008gi}. Using the homological technique of Section \bref{sec:red} one can explicitly show that the reduced BRST charge reads 
\be
\label{lcbrst}
\tilde \brst  = c_0 (2p^+ p_-+p^mp_m)\;,
\ee
so that the equations of motion reduce to the mass-shell condition $p^2 = 0$. 

The light-cone off-shell constraints following from \eqref{BRSTtr}--\eqref{conex} read
\be
\label{lctr}
(\tilde T+\nu)\phi = 0\;, 
\qquad
\tilde T^{\alpha}\phi = 0\;,
\qquad
\tilde T^{\alpha\beta}\phi = 0\;,
\ee
the Young symmetry and spin weight conditions 
\be
\label{ext2}
\tilde N_\alpha{}^\beta \phi = 0 \quad \alpha<\beta\;,  \qquad \tilde N_\alpha\phi =  s_\alpha\phi\;,
\qquad \alpha,\beta = 1, \ldots, n-1\;,
\ee
where 
\be
\label{def2}
\tilde T^{ij} = \frac{\partial^2}{\partial a_i^m \partial a_j{}_m}\;,
\qquad
\tilde N_\alpha{}^\beta = a^m_\alpha \dl{a^\beta{}^m}\;,
\qquad
\tilde N_\alpha = a_\alpha^m\dl{a_\alpha^m}\;,
\ee
no sum over $\alpha$ in $\tilde N_\alpha$. 
The light-cone BRST operator \eqref{lcbrst} obviously acts in the subspace.

\paragraph{Poincare algebra.} The Poincare generators \eqref{poincare} in the light-cone basis split into two groups:  kinematical $G_{kin} = (P^+, P^m, M^{+m}, M^{+-}, M^{mk})$ and dynamical $G_{dyn} = (P^-, M^{-k})$. After quartet reduction both types of generators act in the subspace, $\tilde G_{kin}$ and $\tilde G_{dyn}$.  We find out that the reduced kinematical generators $\tilde G_{kin}$ take the standard form, while the reduced dynamical generators $\tilde G_{dyn}$ are given by 
\be
\label{M-m}
\tilde P^- = - \frac{p^k p_k}{2p^+}\;, 
\qquad
\tilde M^{-m} =  -\frac{\partial}{\partial p^+}p^m - \frac{\partial}{\partial p_m}\frac{p^k p_k}{2p^+}+ \frac{1}{p^+}(S^{mk}p_k+H^m) \;,
\ee 
where   $S^{mn}$ and $H^m$ read
\be
\label{45}
S^{mn} = a_\alpha^m \frac{\partial}{\partial a^\alpha_n} + a^m \frac{\partial}{\partial a_n}-(m \leftrightarrow n)\;,
\qquad\;\;
H_n = \mu \dl{a^n}\;. 
\ee
The elements $S^{kl}$ and $H^n$ satisfy the $iso(d-2)$ commutation relations  
\be
\label{iso}
[S^{kl}, S^{ps}] = \delta^{kp}S^{ls}+ \text{3 terms}\;, 
\qquad
[S^{kl}, H^n] = \delta^{kn}H^l -\delta^{ln}H^k\;, 
\qquad
[H^k, H^l] = 0\;. 
\ee
Note that this oscillator realization of $iso(d-2)$ algebra is analogous to the realization of $iso(d-1,1)$ employed in \cite{Alkalaev:2008gi} in describing mixed-symmetry helicity fields on Minkowski space.

\paragraph{Casimir operators.} To characterize the light-cone representation with given weights we  compute the Casimir operators of the $iso(d-2)$ algebra \eqref{iso}, cf. discussion in Section \bref{sec:C4}. Using \eqref{45} we immediately see that the second order $iso(d-2)$ Casimir operator is given by  
\be
\label{c2}
c_2 \equiv  H^2 \approx  \mu^2\nu\;,
\ee 
where we denoted $H^2 \equiv H_mH^m $ and used the modified trace constraint \eqref{lctr}, cf. \eqref{onshellcas}.  

Using both the (modified) trace conditions \eqref{lctr} and Young and spin weight conditions \eqref{ext2} we explicitly compute the quartic Casimir operator 
\be
\label{c4}
c_4 \equiv H^2 S^2 -2(HS)^2 \approx \mu^2 \nu \sum_{\alpha=1}^{n-1} s_\alpha (s_\alpha +d -2\alpha -3)\;, 
\ee  
where $H^2 = H^m H_m$, $S^2 = S_{mn}S^{mn}$,  $(HS)^m = H_n S^{nm}$. Higher order Casimir operators can be found analogously (see the general discussion in Appendix \bref{sec:casimir}). 

Note that evaluating Casimir operators \eqref{c2} and \eqref{c4} we do not need to solve the constraints \eqref{lctr}--\eqref{ext2} explicitly. The Poincare algebra generators projected onto the subspace singled out by the off-shell constraints are quite complicated. In particular, $iso(d-2)$ subalgebra \eqref{45} is now differently realized: $S^{mn}$ still act as rotations, while translations $H^m$ are non-trivially projected onto the subspace.\footnote{In the spin-$0$ case the explicit realization can be found in \cite{Metsaev:2017cuz} and in the spin-$s$ case in $d=5$ in \cite{Metsaev:2017myp}.}

\subsection{Spin-$s$ case}
\label{sec:spins}

Let us analyze the continuous spin representation labeled by  $(s,0,...,0)$ in more detail. In this case there are two oscillators $(a, a_1^m)$ and the trace constraints read
\be
\label{case}
(\tilde T+\nu)\phi = 0\;, 
\qquad
\tilde T^{1}\phi = 0\;,
\qquad
\tilde T^{11}\phi = 0\;,
\ee
where 
\be
\label{two}
\phi = \sum_{p = 0}^\infty \phi_{m_1 ... m_p|n_1 ... n_s}\; a^{m_1} \cdots a^{m_p}a_1^{n_1} \cdots a_1^{n_s}\;,
\ee
and the spin weight condition $\tilde N_1 \phi = s\phi$ has been taken into account.

Let $Y(k,l)$ denote a traceless $o(d-2)$ tensor associated to the Young diagram with $k$ indices in the first row and $l$ indices in the second row. Then, the solution to  \eqref{case} is given by 
\be
\label{space0}
\phi \;:\qquad  \bigoplus_{l=0}^s \bigoplus_{k=s}^\infty Y(k,l)\;.  
\ee   
It is obtained by tensoring two independent traceless rank-$p$ and rank-$s$ tensors in \eqref{two} and subtracting  cross-traces. 
  
When $s=0$ the space \eqref{space0} is an infinite chain  of totally symmetric $o(d-2)$ traceless tensors \cite{Schuster:2014hca,Metsaev:2016lhs,Metsaev:2017cuz}. In this case,  \eqref{space0} directly follows from the $o(d-2)$ version of Proposition \bref{prop1}. For  $s\neq 0$ the space \eqref{space0} is a light-cone version of the covariant formulation discussed in \cite{Zinoviev:2017rnj}. 

In particular, let us consider $d=5$. Using the fact that $o(3)$ traceless tensors satisfy the Hodge duality relations $Y(k,1) \sim Y(k,0)$ and $Y(k,m) = 0$ at $m>1$ we find out that in this case \eqref{space0} is the representation space described in \cite{Brink:2002zx} (see also \cite{Metsaev:2017myp}), i.e. two infinite chains of traceless $o(3)$ tensors $Y(k,0)$ with $k = s, s+1, ..., \infty$. The quartic Casimir operator \eqref{c4} in five dimensions 
can be represented as $c_4 = W^2$, where  $W = \epsilon_{klm}H^k S^{lm}$. Thus, we reproduce   the result of \cite{Brink:2002zx} that $W = \pm \,\mu\, s$.

\subsection{Mixed-symmetry case}

Here we describe the space of light-cone continuous spin fields for arbitrary spins $(s_1, ..., s_{n})$. Let $Y(l_1, ..., l_p)$ denote a traceless $o(d-2)$ tensor with indices described by Young diagram with $p$ rows of lengths $l_j$, $j=1,...,p$.  Then, the solution to the algebraic constraints \eqref{lctr} and \eqref{ext2}  is given by 
\be
\label{space1}
\phi \;:\qquad  \bigoplus_{k=s_1}^\infty \;\bigoplus_{l_\alpha \leq s_\alpha}\;\;  Y(k,l_1,...,l_{n-1})\;.  
\ee   
In other words, the solution space is given by a finite collection of infinite chains of Young diagrams with the length of the first row running from $s_1$ to infinity. The chains differ from each other by the form of the lower part of  diagrams which is defined by all admissible (consistent with the Young symmetry conditions) lengths $l_\alpha=0,1,..., s_\alpha$.

\section{Weyl and gauge modules}\label{sec:modules}
\label{sec:module}

An important invariant information about a given linear gauge system is encoded in the space of gauge inequivalent formal solutions to the equations of motion, known as Weyl module, and the space of (higher-order) global reducibility parameters, known as a gauge module. These spaces are typically seen as modules over the space-time global symmetry algebra.  In particular, if the gauge module vanishes the system is non-gauge, i.e. all the gauge symmetries are  Stueckelberg-like. Note also, that if the gauge module vanishes and the space-time global symmetries (e.g. Poincare or AdS) act transitively, the system is entirely determined by the Weyl module structure. This property is manifest in the unfolded approach.\footnote{For a  review of the unfolded approach, see e.g.~\cite{Bekaert:2005vh}. Note that the unfolded description for continuous spin fields was implicitly  discussed  in~\cite{Ponomarev:2010st}. Within the present framework, more details on the gauge and Weyl modules can be found in~\cite{Barnich:2015tma,Chekmenev:2015kzf} and references therein.} 

We are now interested in the gauge and Weyl modules of the continuous spin system. To analyze formal solutions in this section we replace space-time coordinates $x^a$ by formal  coordinates $y^a$. In particular, in this section it is implicitly assumed that in all the expressions for fields, parameters, operators, etc. $x^a$ is replaced with $y^a$.  The gauge and Weyl modules can be defined as the cohomology $H^k(Q)$ of the BRST operator 
\be
\label{fiveQ}
Q = \left(\sd_i+\mu_i\right) \dl{b_i}\;, \qquad \text{where}\qquad \sd_i = a^i_a\dl{y^a}\;,
\qquad  i = 1,...,n\;,
\ee 
acting in the formal series singled out by the modified trace and Young symmetry and spin weight constraints \eqref{diff}--\eqref{young}. The Weyl module is the zero ghost number cohomology $H^0(Q)$, the gauge module is a collection of modules identified with negative ghost degree cohomology $H^k(Q)$ at $k<0$ \cite{Alkalaev:2008gi,Alkalaev:2009vm,Alkalaev:2011zv}. 

\subsection{Scalar continuous spin fields} 

To begin with we consider the scalar continuous system ($n=1$) taking  as a functional class formal series in formal variable $y^a$ and auxiliary variable  $a^b$. To compute the cohomology  we first reduce the problem to the  subspace 
\begin{equation}
\label{def-tr}
(T+\nu)\phi=\Box\phi=D \phi=0\,.
\end{equation} 
The Weyl module is then the quotient of this space modulo the image of $\sd+\mu$ (i.e. its cokernel), while the gauge module is the kernel of $\sd+\mu$. 

Let us first characterize the subspace \eqref{def-tr} differently. We have
\begin{prop}\label{prop:trace}
Let $\phi_0(y,a)$ be totally traceless, i.e. 
\be
\label{traceless}
T \phi_0=\Box\phi_0=D \phi_0=0\;.
\ee 
Then, there exists a unique $\phi$ such that $\Pi \phi=\phi_0$, where $\phi$ satisfies \eqref{def-tr},  and $\Pi$ denotes the projector to the totally traceless component. This gives an isomorphism between the space~\eqref{def-tr} and the space of totally traceless elements \eqref{traceless}.
\end{prop}
The proof makes use of the cohomology statements from~\cite{Barnich:2004cr}. The present proposition is analogous to the Proposition \bref{prop1}.

Using the above isomorphism the action of $\sd+\mu$ can be written in terms of totally traceless $\phi_0$ so that the problem is reformulated as that of computing kernel and cokernel of $\sd+\mu$ in the space \eqref{traceless}. 

\begin{prop}
\label{kernelcokernel}
In the space of totally traceless formal series in $y,a$ we have 
\be
\ker(\sd+\mu)=0\;, \qquad \mathrm{coker}(\sd+\mu)=0\;.
\ee
\end{prop}
The first part is trivial. The second is equivalent to the fact that any $\phi_0$ can be represented as $(\sd+\mu)\chi$. In the space of formal series $\chi=(\sd+\mu)^{-1}\phi_0$, where the inverse gauge generator is the Neumann series  
\be
\label{inverse}
\frac{1}{\sd+\mu}\;=\;\mu^{-1}-\mu^{-2}\,\sd +\mu^{-3}\,\sd\sd-\mu^{-4}\,\sd\sd\sd+\ldots\;. 
\ee

It follows from the above statement that formal power series in $y^a$ and $a^b$ is not a satisfactory functional class as the system is empty with such a choice. Although the gauge module $\ker(\sd+\mu)$ remains trivial even if one restricts to a subspace of formal series, the Weyl module can be made nontrivial if in Proposition~\bref{kernelcokernel} one restricts to polynomials in $a$ with coefficients in formal series in $y$ (i.e. $\phi_0$ belongs to the natural class employed when studying the Fronsdal system). 

This can be achieved if before reducing to totally traceless elements one restricts to the following class: formal series in $y^a$ and $a^b$ such that $\Pi \phi$ is a polynomial in $a^b$ with coefficients in formal series in $y^a$.  Indeed, with this choice  Proposition~\bref{prop:trace} remains correct as the space of solutions of~\eqref{def-tr} in this functional space is isomorphic to totally traceless  elements which are polynomials in $a^b$ with coefficients in formal series in $y^a$. 

Getting back to the Weyl module it is straightforward to see that it becomes nontrivial. Indeed, the formal inverse~\eqref{inverse} of $\sd+\mu$ does not  preserve polynomials in $a^b$ as it contains an infinite series in $a^b$. On the other hand,  any polynomial in $y^a$ is in the image of $\sd+\mu$ because the series terminates in this case.

Although it is not clear how to characterize the Weyl module explicitly let us give an example of a nontrivial element. Let $k^a$ be a light-like constant vector, i.e. $k^ak_a=0$. Consider the ``formal plane wave''
\begin{equation}
\label{plane}
 \phi_0=exp(ik_a y^a)\,: \qquad T \phi_0=\Box\phi_0=D \phi_0=0\,.
\end{equation} 
It is a polynomial in $a$ (zero degree), while $(\sd+\mu)^{-1}\phi_0$ is a formal series in $a$. We conclude that the element \eqref{plane} does not belong to the functional class and, therefore, defines a nontrivial cohomology.

Finally, to arrive at the functional class introduced from the very beginning in Section~\bref{sec:cons} one observes that one can equivalently require only the traceless (in $a$-oscillators) component of $\phi$ to be polynomial in $a^b$. Also, for the subspace to be closed with respect to all the $sp(2n+2)$ generators one should require all traces to be polynomial in $a^b$. As we  shown in Section~\ref{sec:light} this is also necessary for the consistent light-cone reduction of the system. The above considerations motivate the functional class choice made in Section~\bref{sec:cons}.

\subsection{Mixed-symmetry continuous spin fields} 

Let us now turn to the general case $n\geq 1$. To compute the cohomology $H(Q)$ we have the following generalization of the Proposition~\bref{prop:trace}.
\begin{prop}\label{prop:trace-gen}
Let $\phi_0(y,a)$ be totally traceless, i.e. $T^{ij} \phi_0 = \Box\phi_0 = D^i \phi_0 = 0$. Then, there exists a unique $\phi$ such that
\begin{equation}
\label{sunday}
 \Pi \phi=\phi_0\,, \qquad (T+\nu)\phi=T^{\alpha}\phi=T^{\alpha\beta}\phi=\Box\phi=D^i\phi=0\,,
\end{equation} 
where $\Pi$ denotes the projector to the totally traceless component. This gives an isomorphism between the space
$(T+\nu)\phi=T^{\alpha}\phi=T^{\alpha\beta}\phi=\Box\phi=D^i\phi=0$ and the space of totally traceless elements.
\end{prop}
The proof is analogous to that of Proposition \bref{prop:trace} and involves a cohomology statement from~\cite{Alkalaev:2008gi}. 

Using the isomorphism \eqref{sunday} one can reformulate the problem in terms of totally traceless elements. Let us explicitly consider the first nontrivial case $n=2$, where only two oscillators $a,a_1$ are present.  The traceless general element $\Phi = \phi + \psi b +\psi^1 b_1 +\chi b b_1$ satisfies BRST extended spin condition \eqref{ext1} and belongs to the functional class defined in Section~\bref{sec:cons}.

{\bf Gauge module.} In degree $-2$ one gets the cocylce condition
\begin{equation}
(\sd+\mu)\chi=0\,, \qquad \sd_1\chi=0 \;,
\end{equation} 
so that the cohomology is empty, cf. \eqref{inverse}.

Now, in degree $-1$ the cocycle and coboundary conditions read as 
\begin{equation}
 (\sd+\mu)\psi+\sd_1\psi^1=0\,,\qquad \psi \sim \psi + \sd_1\lambda\,, \quad \psi^1 \sim \psi^1 -(\sd+\mu)\lambda\;,
\end{equation} 
where degree $-2$ parameter is $\lambda b b_1$. Suppose we are given with a nontrivial solution $\psi b +\psi^1 b_1$. It follows that 
\begin{equation}
\psi=-(\sd+\mu)^{-1}\sd_1\psi^1\;,
\end{equation}
and, hence, there exists $\ell$ such that $(\sd)^\ell(\sd_1\psi^1)=0$ (otherwise $\psi$ is an infinite series in $a$). This in turn implies that $\psi_1$ is polynomial in $y$. Being polynomial it can always be represented as $(\sd+\mu)\kappa$ for some $\kappa$ and, hence, one can assume $\psi^1=0$. It then follows  that $(\sd+\mu)\psi=0$ and, hence, $\psi=0$ as $\ker(\sd+\mu)=0$. We conclude that the cohomology is empty.

{\bf Weyl module.} In degree $0$ the cocycle condition is trivial, while the coboundary condition gives
\begin{equation}
 \phi \sim \phi + (\sd+\mu)\xi+\sd_1\xi^1\;,
\end{equation} 
where degree $-1$ parameter is $\xi b +\xi^1 b_1$. Had we taken as a functional class all formal series in $a$-oscillators, Proposition \bref{kernelcokernel} would have implied that the cohomology is empty. However, using the functional class introduced in Section~\bref{sec:cons} the Weyl module is non-vanishing similarly to the scalar case. Indeed, the following analog of \eqref{plane} gives an example of a nontrivial element of the Weyl module:
\begin{equation}
\phi_0=\Pi\left( g^{s_1}(a_1) exp(ik_a y^a)\right)\;,
\end{equation}
where $g^{s_1}(a_1)$ is a degree $s_1$ polynomial in $a_1$ and $\Pi$ denotes a projector to totally traceless component. Similar reasoning show that $\phi_0$ is nontrivial in cohomology for $g^{s_1}(a_1)$ of general position.

The extension of the above statements to the general mixed-symmetry case is straightforward. We conclude that with the properly chosen functional class for spin oscillators, generic massless continuous spin fields are not genuine gauge fields. In particular, the gauge fields present in various formulations of continuous spin systems should be Stueckelberg fields. At the same time the Weyl module is nontrivial in agreement with the light-cone formulation of Section \bref{sec:light}.
 
\section{Conclusion}

In this paper we developed the BRST-based approach to  continuous spin fields of arbitrary mixed-symmetry type in Minkowski space. Using the Howe duality between $o(d-1,1)$ Lorentz algebra and $sp(2n+2)$ symplectic algebra we formulated a set of Poincare invariant  constraints underlying the continuous spin dynamics. The constraint set consists of both the algebraic and differential conditions and can be viewed as the $\mu$-deformation of the helicity spin case, where $\mu$ is the continuous spin parameter.    

Implementing  differential constraints via the BRST operator and imposing  algebraic constraints directly we arrive at the triplet formulation for continuous spin. The resulting equations of motion \eqref{eomtriplet}, \eqref{standard-brst} have a simple form even in the general mixed-symmetry case.  Using the homological reductions of the triplet BRST operator we found the metric-like formulation \eqref{lab} that generalizes the Schuster-Toro description of the scalar continuous spin fields. On the other hand, the resulting metric-like formulation is the $\mu$-deformation of the Labastida equations.    

Applying the so-called quartet mechanism we can get rid of the unphysical components of the oscillators to obtain the light-cone form of the continuous spin dynamics. In particular, we explicitly built the $iso(d-2)$ Wigner little algebra and computed its second and fourth Casimir operators.   

Our formulation can be  naturally extended in several ways. First of all, we completely left aside the Lagrangian formulation for the equations we considered in this paper. Formally, the triplet BRST operator \eqref{standard-brst} is not Hermitian with respect to the standard inner product contrary to the helicity spin case. On the other hand, the Schuster-Toro equations are known to be variational \cite{Schuster:2014hca,Metsaev:2016lhs} and, hopefully, there is a modified inner product that would determine the action functional. Also, the continuous spin fields are known to  consistently propagate on the (A)dS background \cite{Ponomarev:2010st,Metsaev:2016lhs,Zinoviev:2017rnj,Metsaev:2017ytk,Bekaert:2017khg,Khabarov:2017lth}. Thus, the AdS formulation for arbitrary mixed-symmetry continuous spin fields along the lines of \cite{Alkalaev:2009vm,Alkalaev:2011zv} seems to exist though a group-theoretical description of such fields is yet to be found.       
Furthermore, it would be interesting to extend the recent results on interacting scalar continuous fields \cite{Metsaev:2017cuz,Bekaert:2017xin} and elaborate on cubic vertices for mixed-symmetry  fields using either the metric-like formulation \cite{Metsaev:2012uy} or the frame-like formulation \cite{Alkalaev:2010af,Boulanger:2011qt,Zinoviev:2011fv}. Finally, the approach and results of this work should extend smoothly to the case of fermionic continuous spin fields.

\vspace{7mm}
\noindent \textbf{Acknowledgements.} We are grateful to R. Metsaev and D. Ponomarev  for valuable discussions and X. Bekaert for useful comments.  The work of K.A. is supported by the grant RFBR No 17-02-00317. The work of M.G. was supported by the DFG Transregional Collaborative Research Centre TRR 33 and the DFG cluster of excellence ``Origin and Structure of the Universe''.

\appendix

\section{Casimir operators of the Poincare algebra}
\label{sec:casimir}

Let us consider the  Casimir operators  $C_{2p}$, where $ p =1,2,... \;$.  We introduce the generalized Pauli-Lubanski tensors 
\be
W_{m_1 ... m_k} = \epsilon_{m_1 ... m_k a_{k+1} ... a_{d}} P^{a_{k+1}}M^{a_{k+2}a_{k+3}} ... M^{a_{d-1}a_d}\;,
\ee 
where $\epsilon_{m_1 ... m_d}$ is the $o(d-1,1)$ Levi-Civita tensor,  $k = 1,3,..., d-3$ for even $d$ and $k = 0,2,..., d-3$ for odd $d$. The Pauli-Lubanski  tensors  covariantly transform under Lorentz subalgebra $o(d-1,1)$ and satisfy $[P_a, W_{m_1 ... m_k}]=0$ that allows us to represent the Casimir operators as follows 
\be
\label{casimir}
C_{2p} = W_{m_1 ... m_{p-1}}W^{m_1 ... m_{p-1}}\;.
\ee  

Basis elements of the Wigner little algebras can be read off from the tensor $W_{m_1...m_{d-3}}$, which is the direct counterpart of the original $d=4$ Pauli-Lubanski vector. In the massive case we choose the standard momentum representative of the $P^2 = m^2 $ condition  as $P^m = (m,0,...,0)$.  Then,  $W_{m_1...m_{d-3}}$ can be split into components $W_{i_1 ... i_{d-3}} = m \epsilon_{i_1 ... i_{d-1}}M^{i_{d-2}i_{d-1}}$ which are simply the dualized $o(d-1)$ basis elements. The massless case is more intricate. Here, we choose the standard momentum representative $P^m = (\varkappa, 0, ..., 0, \varkappa)$. Then,   $W_{m_1...m_{d-3}}$ similarly splits into $o(d-2)$ rotations and $d-2$ translations that altogether form $iso(d-2)$ algebra.

For arbitrary representations the Casimir operators can be rather complicated, but in the massless case $P^2 = 0$ they are drastically simplified. First of all, we observe that all the Casimir operators are bilinear in momenta,  $C_{2p} \sim F^{mn}(M)P_mP_n$, where $F$ is a polynomial of Lorentz generators. Moreover, one can show that in the massless case momenta  $P_m$ enter only via  combination $M_{ab}P^b$. Denoting $\pi_a = M_{ab}P^b $  we find the general expression  
\be
C_{2p} \approx \left[a_{p,0}+ a_{p,2}\, M^2 + ... + a_{p,2p-4}\, M^{2p-4}\right] \pi_a\pi^a\;,
\ee  
where $\alpha_{p,i}$ are numerical  coefficients, $M^{2k}$ are order $2k$ polynomials in $M_{ab}$, the weak equality means that evaluating \eqref{casimir} we use  the $P^2 = 0$ condition. For example, the quartic Casimir operator \footnote{An explicit expression reads $C_4 = P^2 M^2 - 2 \pi_a\pi^a \approx -2 \pi_a\pi^a$, see e.g. Ref. \cite{Bekaert:2006py}.} is given by $C_4 \sim \pi_a\pi^a$. It is clear then that $C_2 = 0$ defines masslessness, $C_4$ yields the continuous spin value $\mu^2$, while a number of independent spin weights equals to that of the Casimir operators, $C_{6}, C_{8}, ...\;$. In other words, a continuous spin representation is characterized by the parameter $\mu$ and (half-)integers $s_1, ..., s_r$, where $r \leq [\frac{d-2}{2}]-1$. We see that contrary to the helicity spin case  there are one less spin weights.

\section{Trace decompositions}
\label{app:A}

In what follows we solve the deformed trace conditions in the case of totally symmetric tensors.  Let $F = F(a)$ be a generating series
\be
F = \sum_{n=0}^\infty {F_{m_1 ... m_n}} \,a^{m_1} \cdots a^{m_n} \equiv \sum_{n=0}^\infty F_{(n)}\;,
\ee
where $N F_{(n)} = n F_{(n)}$. Let us consider  single or double trace conditions 
\be
\label{trF1}
(T+\nu)F = 0\;, 
\ee
\be
\label{trF2}
(T+\nu)^2 F = 0\;.
\ee
\vspace{-5mm}
\begin{prop}
\label{prop1}
Solutions to  \eqref{trF1} and \eqref{trF2} are respectively given by 
\be
\label{F1}
\quad \qquad \;\; F = \sum_{n,m=0}^\infty \alpha_{m,n} (T^\dagger)^m f_{(n)}\;, \qquad T f_{(n)} = 0\;,
\qquad 
\alpha_{m,n} = \alpha_n\,\frac{ 4^{-m+1} \nu ^{m-1}}{m!\left(\frac{d}{2}+n-1\right)_{m-1}}\;, 
\ee 
and
\be
\label{F2}
F = \sum_{n,m=0}^\infty \beta_{m,n} (T^\dagger)^m f_{(n)}\;, \qquad T^2 f_{(n)} = 0\;,
\qquad
\beta_{m,n} = \beta_n\,\frac{ 4^{-m+1} \nu ^{m-1}}{m! \left(\frac{d}{2}+n\right)_{m-1}}\;, 
\ee
where $\alpha_n = \alpha_n(n,d)$ and $\beta_n = \beta_n(n,d)$ are arbitrary prefactors depending on rank and dimensionality,    and  $(a)_m$ is the Pochhammer symbol. Up to  prefactors the coefficients are related as $\alpha_{m,n} = \beta_{m,n+1}$. 
 
\end{prop}
To prove the proposition we reformulate the deformed trace conditions as recursive equations on the coefficients $\alpha_{m,n}$ and $\beta_{m,n}$. The general solution is parameterized by $n$-dependent constants. Let us note that the solution exists when e.g. $\alpha_n =0$ at $\forall n \neq n_0$.

\section{Reproducing the Schuster-Toro equations}
\label{app:shto}

Let us consider first the deformed gauge transformation \eqref{deform}. In the Fronsdal basis \eqref{F1}, \eqref{F2} $n$-th rank Fronsdal field is transformed as 
\be
\label{saturday}
\delta \varphi_{(n)} = D^\dagger \epsilon_{(n-1)} + \rho_n\, \, \epsilon_{(n)} + \gamma_n \,T^\dagger \epsilon_{(n-2)}\;,  
\ee   
where the coefficients are proportional to the continuous spin parameter $\mu$. To find them we have to solve the system of recurrent equations coming from representing  both sides of \eqref{deform} in the Fronsdal basis. Fixing the number $\# a_i^\mu = p$ we find that 
\be
\label{saturday2}
\ba{r}
\dps
\beta_{0,p} \delta\varphi_{(p)} = \sum_{k=0}^{\frac{p-1}{2}} \alpha_{k, p-2k-1}(T^\dagger)^k \epsilon_{(p-2k-1)}+ \mu \sum_{k=0}^{\frac{p}{2}} \alpha_{k, p-2k}(T^\dagger)^k D^\dagger\epsilon_{(p-2k)}-
\\
\\
\dps
-\sum_{k=0}^{\frac{p-1}{2}} \alpha_{k, p-2k}(T^\dagger)^k D^\dagger\delta\varphi_{(p-2k)}\;.
\ea
\ee  
Substituting \eqref{saturday} into \eqref{saturday2} and using the property $\alpha_{n,s} = \beta_{n,s+1}$ (see Proposition \bref{prop1}) we find that in the $p$-th order  
\be
\label{d1}
\rho_p = \frac{\alpha_{0,p}}{\beta_{0,p}}\;,
\qquad
\gamma_p = \frac{\alpha_{1,p-2}-\beta_{1,p-2}\rho_{p-2}}{\beta_{0,p}}\;,
\ee 
while in the lower orders we find the system 
\be
\label{d2}
\alpha_{k,p-2k}-\beta_{k,p-2k}\rho_{p-2k} - \beta_{k-1,p-2k+2}\gamma_{p-2k+2} = 0\;,
\qquad  k=1,2,...,\frac{p}{2}\;.
\ee   
Fixing  normalization constants in \eqref{F2} as $\beta_n=\frac{d}{d+2 n-2}$ we find that
\be
\rho_n = \mu\;, \qquad \gamma_n = -\frac{\mu\nu}{(d+2n-4)(d+2n-6)}\,,
\ee
cf. \eqref{shto_trans}, and that the equation system \eqref{d2} is identically satisfied. 

The equations of motion \eqref{red_eom} can be analyzed along the same lines. Substituting the Fronsdal decomposition \eqref{trd} into \eqref{red_eom} we find that the equations take the form 
\be
\label{redF}
\ba{l}
\dps
\;\;\;\sum_{s,n=0}^\infty \beta_{n,s} (T^\dagger)^n F_{(s)} +\sum_{s,n=0}^\infty \zeta_{n,s} (T^\dagger)^n D \varphi_{(s)}+\sum_{s,n=0}^\infty \gamma_{n,s} (T^\dagger)^n D^\dagger \varphi_{(s)}+
\\
\\
\dps
+\sum_{s,n=0}^\infty \rho_{n,s} (T^\dagger)^n D^\dagger T \varphi_{(s)}+
\sum_{s,n=0}^\infty \tau_{n,s} (T^\dagger)^n \varphi_{(s)}+\sum_{s,n=0}^\infty \varkappa_{n,s} (T^\dagger)^n T \varphi_{(s)} = 0\;,
\ea
\ee    
where 
\be
F_{(n)} = \left[\Box  -D^{\dagger}  D   + \half D^{\dagger} D^{\dagger}   T\right]\varphi_{(n)}\;,
\ee
is the standard Fronsdal kinetic operator, the coefficients are given by 
\be
\ba{c}
\dps
\zeta_{n,s} = -\mu \beta_{n,s} \;,
\qquad
\gamma_{n,s}  = 2 \mu (n+1)\beta_{n+1,s}\;,
\qquad
\rho_{n,s} = \mu \beta_{n,s}\;,
\\
\\
\dps
\tau_{n,s} = 2 \mu^2 (n+1)\beta_{n+1,s}\;,
\qquad 
\varkappa_{n,s} =\half \mu^2 \beta_{n,s}\;. 
\ea
\ee 
We want to represent \eqref{redF} as 
\be
\label{EEEE}
\sum_{n,m=0}^\infty \beta_{m,n} (T^\dagger)^m \left[F_{(n)}+ E_{(n)}\right] = 0\;, 
\ee
so that the equations for the $n$-th rank contribution take the form $F_{(n)}+ E_{(n)} = 0$, where the $\mu$-correction is double traceless, $T^2 E_{(n)} = 0$. The form of \eqref{redF} suggests the following general expression
\be
\label{Ep}
\ba{r}
\dps
E_{(p)} = a_p  D   \varphi_{(p+1)}+ b_p D^\dagger T \varphi_{(p+1)}+c_p D^\dagger \varphi_{(p-1)}+d_p D^\dagger T^\dagger T \varphi_{(p-1)}+ e_p T^\dagger  D   \varphi_{(p-1)}+
\\
\\
+\bar a_p  \varphi_{(p)}+ \bar b_p T^\dagger \varphi_{(p-2)}+\bar c_p T \varphi_{(p+2)}+\bar d_p T^\dagger T \varphi_{(p)}+ \bar e_p T^\dagger T^\dagger \varphi_{(p-4)}\;,
\dps	
\ea
\ee
where coefficients are not independent and related by the double trace condition imposed both on $\varphi_{(k)}$ and $E_{(l)}$. Now, let us introduce  particular traceless combinations, cf. \eqref{AB},
\be
\label{AB1}
A_{(n)} =  D   \varphi_{(n+1)} - \half D^\dagger\, T \varphi_{(n+1)}\;,
\qquad 
B_{(n)} = \varphi_{(n)} + y_n\, T^\dagger T  \varphi_{(n)}+ z_n\, T \varphi_{(n+2)}\;,
\ee
where the coefficients are fixed by the zero trace condition 
as $y_n = - (2d+2n-8)^{-1}$ and $\forall z_n$. Then, we can represent \eqref{Ep} as
\be
\label{EEE}
E_{(p)} = \tilde \varkappa_{p} A_{(p)} +\tilde \tau_{p} T^\dagger A_{(p-2)}+\tilde \rho_{p} D^\dagger B_{(p-1)} + \bar \rho_{p} B_{(p)} + \bar \varkappa_p T^\dagger B_{(p-2)} \;,
\ee
while the Fronsdal operator takes the form $F_{(n)} = \Box\varphi_{(p)} -D^\dagger A_{(p-1)}$. The coefficients  $\tilde \varkappa_{p}, \tilde \tau_{p}, \tilde \rho_{p}, \bar \rho_{p}, \bar \varkappa_p$ and $z_n$ are expressed via $a_p, b_p, c_p, d_p, e_p$ and  $\bar a_p, \bar b_p, \bar c_p, \bar d_p, \bar e_p$ and $y_n$. Substituting \eqref{EEE} into \eqref{EEEE} and performing the analysis similar to what we did for the gauge transformations we find that coefficients in \eqref{EEE} are fixed as follows
\be
\ba{c}
\dps
\tilde \varkappa_{p} = -\mu\;, 
\qquad
\tilde \tau_{p} = \frac{\mu \nu }{(d+2p-4)(d+2p-6)}\;,
\qquad
\tilde \rho_{p} = \frac{\mu\nu}{d+2p-4}\;,
\\
\\
\dps
\bar \rho_p = \frac{\mu^2 \nu}{d+2p-2}\;, 
\qquad
\bar \varkappa_p = \frac{\mu^2 \nu^2 }{(d+2p-4)(d+2p-6)^2}\;,
\qquad
z_p = \frac{d+2p-2}{2\nu}\;.
\ea
\ee   
Reorganizing the equation \eqref{EEE} and the Fronsdal operator  in terms of $G_{(p)}$ \eqref{Gn} we finally arrive at the Schuster-Toro equations \eqref{shto}. 

\providecommand{\href}[2]{#2}\begingroup\raggedright\endgroup


\begin{thebibliography}{10}

\bibitem{Bargmann:1948ck}
V.~Bargmann and E.~P. Wigner, \emph{{Group Theoretical Discussion of
  Relativistic Wave Equations}},
  \href{http://dx.doi.org/10.1073/pnas.34.5.211}{\emph{Proc. Nat. Acad. Sci.}
  {\bf 34} (1948) 211}.

\bibitem{Bekaert:2005in}
X.~Bekaert and J.~Mourad, \emph{{The Continuous spin limit of higher spin field
  equations}},
  \href{http://dx.doi.org/10.1088/1126-6708/2006/01/115}{\emph{JHEP} {\bf 01}
  (2006) 115}, [\href{http://arxiv.org/abs/hep-th/0509092}{{\tt
  hep-th/0509092}}].

\bibitem{Bengtsson:2013vra}
A.~K.~H. Bengtsson, \emph{{BRST Theory for Continuous Spin}},
  \href{http://dx.doi.org/10.1007/JHEP10(2013)108}{\emph{JHEP} {\bf 10} (2013)
  108}, [\href{http://arxiv.org/abs/1303.3799}{{\tt 1303.3799}}].

\bibitem{Schuster:2014hca}
P.~Schuster and N.~Toro, \emph{{Continuous-spin particle field theory with
  helicity correspondence}},
  \href{http://dx.doi.org/10.1103/PhysRevD.91.025023}{\emph{Phys. Rev.} {\bf
  D91} (2015) 025023}, [\href{http://arxiv.org/abs/1404.0675}{{\tt
  1404.0675}}].

\bibitem{Rivelles:2014fsa}
V.~O. Rivelles, \emph{{Gauge Theory Formulations for Continuous and Higher Spin
  Fields}}, \href{http://dx.doi.org/10.1103/PhysRevD.91.125035}{\emph{Phys.
  Rev.} {\bf D91} (2015) 125035}, [\href{http://arxiv.org/abs/1408.3576}{{\tt
  1408.3576}}].

\bibitem{Najafizadeh:2015uxa}
X.~Bekaert, M.~Najafizadeh and M.~R. Setare, \emph{{A gauge field theory of
  fermionic Continuous-Spin Particles}},
  \href{http://dx.doi.org/10.1016/j.physletb.2016.07.005}{\emph{Phys. Lett.}
  {\bf B760} (2016) 320--323}, [\href{http://arxiv.org/abs/1506.00973}{{\tt
  1506.00973}}].

\bibitem{Metsaev:2016lhs}
R.~R. Metsaev, \emph{{Continuous spin gauge field in (A)dS space}},
  \href{http://dx.doi.org/10.1016/j.physletb.2017.02.027}{\emph{Phys. Lett.}
  {\bf B767} (2017) 458--464}, [\href{http://arxiv.org/abs/1610.00657}{{\tt
  1610.00657}}].

\bibitem{Metsaev:2017ytk}
R.~R. Metsaev, \emph{{Fermionic continuous spin gauge field in (A)dS space}},
  \href{http://arxiv.org/abs/1703.05780}{{\tt 1703.05780}}.

\bibitem{Zinoviev:2017rnj}
{\relax Yu}.~M. Zinoviev, \emph{{Infinite spin fields in d = 3 and beyond}},
  \href{http://dx.doi.org/10.3390/universe3030063}{\emph{Universe} {\bf 3}
  (2017) 63}, [\href{http://arxiv.org/abs/1707.08832}{{\tt 1707.08832}}].

\bibitem{Najafizadeh:2017tin}
M.~Najafizadeh, \emph{{Modified Wigner equations and continuous spin gauge
  field}},  \href{http://arxiv.org/abs/1708.00827}{{\tt 1708.00827}}.

\bibitem{Bekaert:2017khg}
X.~Bekaert and E.~D. Skvortsov, \emph{{Elementary particles with continuous
  spin}}, \href{http://dx.doi.org/10.1142/S0217751X17300198}{\emph{Int. J. Mod.
  Phys.} {\bf A32} (2017) 1730019},
  [\href{http://arxiv.org/abs/1708.01030}{{\tt 1708.01030}}].

\bibitem{Rehren:2017xzn}
K.-H. Rehren, \emph{{Pauli-Lubanski limit and stress-energy tensor for
  infinite-spin fields}},  \href{http://arxiv.org/abs/1709.04858}{{\tt
  1709.04858}}.

\bibitem{Metsaev:2017cuz}
R.~R. Metsaev, \emph{{Cubic interaction vertices for continuous-spin fields and
  arbitrary spin massive fields}},  \href{http://arxiv.org/abs/1709.08596}{{\tt
  1709.08596}}.

\bibitem{Bekaert:2017xin}
X.~Bekaert, J.~Mourad and M.~Najafizadeh, \emph{{Continuous-spin field
  propagator and interaction with matter}},
  \href{http://arxiv.org/abs/1710.05788}{{\tt 1710.05788}}.

\bibitem{Khabarov:2017lth}
M.~V. Khabarov and {\relax Yu}.~M. Zinoviev, \emph{{Infinite (continuous) spin
  fields in the frame-like formalism}},
  \href{http://dx.doi.org/10.1016/j.nuclphysb.2018.01.016}{\emph{Nucl. Phys.}
  {\bf B928} (2018) 182--216}, [\href{http://arxiv.org/abs/1711.08223}{{\tt
  1711.08223}}].

\bibitem{Metsaev:2017myp}
R.~R. Metsaev, \emph{{Continuous-spin mixed-symmetry fields in AdS(5)}},
  \href{http://arxiv.org/abs/1711.11007}{{\tt 1711.11007}}.

\bibitem{Vasiliev:2003ev}
M.~A. Vasiliev, \emph{{N}onlinear equations for symmetric massless higher spin
  fields in ({A})d{S}(d)}, {\emph{Phys. Lett.} {\bf B567} (2003) 139--151},
  [\href{http://arxiv.org/abs/hep-th/0304049}{{\tt hep-th/0304049}}].

\bibitem{Bekaert:2005vh}
X.~Bekaert, S.~Cnockaert, C.~Iazeolla and M.~A. Vasiliev, \emph{{Nonlinear
  higher spin theories in various dimensions}},
  \href{http://arxiv.org/abs/hep-th/0503128}{{\tt hep-th/0503128}}.

\bibitem{Brink:2002zx}
L.~Brink, A.~M. Khan, P.~Ramond and X.-z. Xiong, \emph{{Continuous spin
  representations of the Poincare and superPoincare groups}},
  \href{http://dx.doi.org/10.1063/1.1518138}{\emph{J. Math. Phys.} {\bf 43}
  (2002) 6279}, [\href{http://arxiv.org/abs/hep-th/0205145}{{\tt
  hep-th/0205145}}].

\bibitem{Bekaert:2006py}
X.~Bekaert and N.~Boulanger, \emph{{The Unitary representations of the Poincare
  group in any spacetime dimension}},  in \emph{{2nd Modave Summer School in
  Theoretical Physics Modave, Belgium, August 6-12, 2006}}, 2006.
\newblock \href{http://arxiv.org/abs/hep-th/0611263}{{\tt hep-th/0611263}}.

\bibitem{Barnich:2004cr}
G.~Barnich, M.~Grigoriev, A.~Semikhatov and I.~Tipunin, \emph{Parent field
  theory and unfolding in {BRST} first-quantized terms}, {\emph{Commun. Math.
  Phys.} {\bf 260} (2005) 147--181},
  [\href{http://arxiv.org/abs/hep-th/0406192}{{\tt hep-th/0406192}}].

\bibitem{Alkalaev:2008gi}
K.~B. Alkalaev, M.~Grigoriev and I.~Y. Tipunin, \emph{{Massless Poincare
  modules and gauge invariant equations}},
  \href{http://dx.doi.org/10.1016/j.nuclphysb.2009.08.007}{\emph{Nucl. Phys.}
  {\bf B823} (2009) 509--545}, [\href{http://arxiv.org/abs/0811.3999}{{\tt
  0811.3999}}].

\bibitem{Alkalaev:2009vm}
K.~B. Alkalaev and M.~Grigoriev, \emph{{Unified BRST description of AdS gauge
  fields}},
  \href{http://dx.doi.org/10.1016/j.nuclphysb.2010.04.004}{\emph{Nucl. Phys.}
  {\bf B835} (2010) 197--220}, [\href{http://arxiv.org/abs/0910.2690}{{\tt
  0910.2690}}].

\bibitem{Alkalaev:2011zv}
K.~Alkalaev and M.~Grigoriev, \emph{{Unified BRST approach to (partially)
  massless and massive AdS fields of arbitrary symmetry type}},
  \href{http://dx.doi.org/10.1016/j.nuclphysb.2011.08.005}{\emph{Nucl. Phys.}
  {\bf B853} (2011) 663--687}, [\href{http://arxiv.org/abs/1105.6111}{{\tt
  1105.6111}}].

\bibitem{Bengtsson:1986ys}
A.~K.~H. Bengtsson, \emph{A unified action for higher spin gauge bosons from
  covariant string theory}, {\emph{Phys. Lett.} {\bf B182} (1986) 321}.

\bibitem{Francia:2002pt}
D.~Francia and A.~Sagnotti, \emph{On the geometry of higher-spin gauge fields},
  {\emph{Class. Quant. Grav.} {\bf 20} (2003) S473--S486},
  [\href{http://arxiv.org/abs/hep-th/0212185}{{\tt hep-th/0212185}}].

\bibitem{Sagnotti:2003qa}
A.~Sagnotti and M.~Tsulaia, \emph{On higher spins and the tensionless limit of
  string theory}, {\emph{Nucl. Phys.} {\bf B682} (2004) 83--116},
  [\href{http://arxiv.org/abs/hep-th/0311257}{{\tt hep-th/0311257}}].

\bibitem{Fronsdal:1978rb}
C.~Fronsdal, \emph{{Massless Fields with Integer Spin}},
  \href{http://dx.doi.org/10.1103/PhysRevD.18.3624}{\emph{Phys. Rev.} {\bf D18}
  (1978) 3624}.

\bibitem{Labastida:1989kw}
J.~M.~F. Labastida, \emph{Massless particles in arbitrary representations of
  the {L}orentz group}, {\emph{Nucl. Phys.} {\bf B322} (1989) 185}.

\bibitem{Lopatin:1988hz}
V.~E. Lopatin and M.~A. Vasiliev, \emph{Free massless bosonic fields of
  arbitrary spin in d- dimensional de {S}itter space}, {\emph{Mod. Phys. Lett.}
  {\bf A3} (1988) 257}.

\bibitem{Buchbinder:2001bs}
I.~L. Buchbinder, A.~Pashnev and M.~Tsulaia, \emph{{Lagrangian formulation of
  the massless higher integer spin fields in the AdS background}},
  \href{http://dx.doi.org/10.1016/S0370-2693(01)01268-0}{\emph{Phys. Lett.}
  {\bf B523} (2001) 338--346}, [\href{http://arxiv.org/abs/hep-th/0109067}{{\tt
  hep-th/0109067}}].

\bibitem{Zinoviev:2001dt}
Y.~M. Zinoviev, \emph{{On massive high spin particles in (A)dS}},
  \href{http://arxiv.org/abs/hep-th/0108192}{{\tt hep-th/0108192}}.

\bibitem{Alkalaev:2003qv}
K.~B. Alkalaev, O.~V. Shaynkman and M.~A. Vasiliev, \emph{On the frame-like
  formulation of mixed-symmetry massless fields in {(A)dS(d)}}, {\emph{Nucl.
  Phys.} {\bf B692} (2004) 363--393},
  [\href{http://arxiv.org/abs/hep-th/0311164}{{\tt hep-th/0311164}}].

\bibitem{Skvortsov:2008sh}
E.~D. Skvortsov, \emph{{Frame-like Actions for Massless Mixed-Symmetry Fields
  in Minkowski space}},  \href{http://arxiv.org/abs/0807.0903}{{\tt
  0807.0903}}.

\bibitem{Campoleoni:2008jq}
A.~Campoleoni, D.~Francia, J.~Mourad and A.~Sagnotti, \emph{{Unconstrained
  Higher Spins of Mixed Symmetry. I. Bose Fields}},
  \href{http://arxiv.org/abs/0810.4350}{{\tt 0810.4350}}.

\bibitem{Skvortsov:2009zu}
E.~D. Skvortsov, \emph{{Gauge fields in (anti)-de Sitter space and Connections
  of its symmetry algebra}},  \href{http://arxiv.org/abs/0904.2919}{{\tt
  0904.2919}}.

\bibitem{Boulanger2009}
N.~Boulanger, C.~Iazeolla and P.~Sundell, \emph{{Unfolding Mixed-Symmetry
  Fields in AdS and the BMV Conjecture: I. General Formalism}},
  \href{http://dx.doi.org/10.1088/1126-6708/2009/07/013}{\emph{JHEP} {\bf 07}
  (2009) 013}, [\href{http://arxiv.org/abs/0812.3615}{{\tt 0812.3615}}].

\bibitem{Campoleoni:2012th}
A.~Campoleoni and D.~Francia, \emph{{Maxwell-like Lagrangians for higher
  spins}}, \href{http://dx.doi.org/10.1007/JHEP03(2013)168}{\emph{JHEP} {\bf
  03} (2013) 168}, [\href{http://arxiv.org/abs/1206.5877}{{\tt 1206.5877}}].

\bibitem{Francia:2012rg}
D.~Francia, \emph{{Generalised connections and higher-spin equations}},
  \href{http://dx.doi.org/10.1088/0264-9381/29/24/245003}{\emph{Class. Quant.
  Grav.} {\bf 29} (2012) 245003}, [\href{http://arxiv.org/abs/1209.4885}{{\tt
  1209.4885}}].

\bibitem{Bekaert:2015fwa}
X.~Bekaert, N.~Boulanger and D.~Francia, \emph{{Mixed-symmetry multiplets and
  higher-spin curvatures}},
  \href{http://dx.doi.org/10.1088/1751-8113/48/22/225401}{\emph{J. Phys.} {\bf
  A48} (2015) 225401}, [\href{http://arxiv.org/abs/1501.02462}{{\tt
  1501.02462}}].

\bibitem{Joung:2016naf}
E.~Joung and K.~Mkrtchyan, \emph{{Weyl Action of Two-Column Mixed-Symmetry
  Field and Its Factorization Around (A)dS Space}},
  \href{http://dx.doi.org/10.1007/JHEP06(2016)135}{\emph{JHEP} {\bf 06} (2016)
  135}, [\href{http://arxiv.org/abs/1604.05330}{{\tt 1604.05330}}].

\bibitem{Edgren:2006un}
L.~Edgren and R.~Marnelius, \emph{{Covariant quantization of infinite spin
  particle models, and higher order gauge theories}},
  \href{http://dx.doi.org/10.1088/1126-6708/2006/05/018}{\emph{JHEP} {\bf 05}
  (2006) 018}, [\href{http://arxiv.org/abs/hep-th/0602088}{{\tt
  hep-th/0602088}}].

\bibitem{Howe}
R.~Howe, \emph{Transcending classical invariant theory}, {\emph{J. Amer. Math.
  Soc.} {\bf 3} (1989) 2}.

\bibitem{Shaynkman:2000ts}
O.~V. Shaynkman and M.~A. Vasiliev, \emph{{Scalar field in any dimension from
  the higher spin gauge theory perspective}},
  \href{http://dx.doi.org/10.1007/BF02551402}{\emph{Theor. Math. Phys.} {\bf
  123} (2000) 683--700}, [\href{http://arxiv.org/abs/hep-th/0003123}{{\tt
  hep-th/0003123}}].

\bibitem{Henneaux:1990ua}
M.~Henneaux, \emph{Elimination of the auxiliary fields in the antifield
  formalism}, {\emph{Phys. Lett.} {\bf B238} (1990) 299}.

\bibitem{Aisaka:2004ga}
Y.~Aisaka and Y.~Kazama, \emph{{Relating Green-Schwarz and extended pure spinor
  formalisms by similarity transformation}}, {\emph{JHEP} {\bf 04} (2004) 070},
  [\href{http://arxiv.org/abs/hep-th/0404141}{{\tt hep-th/0404141}}].

\bibitem{Barnich:2005ga}
G.~Barnich, G.~Bonelli and M.~Grigoriev, \emph{From {BRST} to light-cone
  description of higher spin gauge fields},
  \href{http://arxiv.org/abs/hep-th/0502232}{{\tt hep-th/0502232}}.

\bibitem{Ponomarev:2010st}
D.~S. Ponomarev and M.~A. Vasiliev, \emph{{Frame-Like Action and Unfolded
  Formulation for Massive Higher-Spin Fields}},
  \href{http://dx.doi.org/10.1016/j.nuclphysb.2010.06.007}{\emph{Nucl. Phys.}
  {\bf B839} (2010) 466--498}, [\href{http://arxiv.org/abs/1001.0062}{{\tt
  1001.0062}}].

\bibitem{Barnich:2015tma}
G.~Barnich, X.~Bekaert and M.~Grigoriev, \emph{{Notes on conformal invariance
  of gauge fields}},
  \href{http://dx.doi.org/10.1088/1751-8113/48/50/505402}{\emph{J. Phys.} {\bf
  A48} (2015) 505402}, [\href{http://arxiv.org/abs/1506.00595}{{\tt
  1506.00595}}].

\bibitem{Chekmenev:2015kzf}
A.~Chekmenev and M.~Grigoriev, \emph{{Boundary values of mixed-symmetry
  massless fields in AdS space}},
  \href{http://dx.doi.org/10.1016/j.nuclphysb.2016.10.006}{\emph{Nucl. Phys.}
  {\bf B913} (2016) 769--791}, [\href{http://arxiv.org/abs/1512.06443}{{\tt
  1512.06443}}].

\bibitem{Metsaev:2012uy}
R.~R. Metsaev, \emph{{BRST-BV approach to cubic interaction vertices for
  massive and massless higher-spin fields}},
  \href{http://dx.doi.org/10.1016/j.physletb.2013.02.009}{\emph{Phys. Lett.}
  {\bf B720} (2013) 237--243}, [\href{http://arxiv.org/abs/1205.3131}{{\tt
  1205.3131}}].

\bibitem{Alkalaev:2010af}
K.~Alkalaev, \emph{{FV-type action for $AdS_5$ mixed-symmetry fields}},
  \href{http://dx.doi.org/10.1007/JHEP03(2011)031}{\emph{JHEP} {\bf 03} (2011)
  031}, [\href{http://arxiv.org/abs/1011.6109}{{\tt 1011.6109}}].

\bibitem{Boulanger:2011qt}
N.~Boulanger, E.~D. Skvortsov and {\relax Yu}.~M. Zinoviev,
  \emph{{Gravitational cubic interactions for a simple mixed-symmetry gauge
  field in AdS and flat backgrounds}},
  \href{http://dx.doi.org/10.1088/1751-8113/44/41/415403}{\emph{J. Phys.} {\bf
  A44} (2011) 415403}, [\href{http://arxiv.org/abs/1107.1872}{{\tt
  1107.1872}}].

\bibitem{Zinoviev:2011fv}
{\relax Yu}.~M. Zinoviev, \emph{{Gravitational cubic interactions for a massive
  mixed symmetry gauge field}},
  \href{http://dx.doi.org/10.1088/0264-9381/29/1/015013}{\emph{Class. Quant.
  Grav.} {\bf 29} (2012) 015013}, [\href{http://arxiv.org/abs/1107.3222}{{\tt
  1107.3222}}].

\end{thebibliography}

\end{document}